# An Interference-Aware Virtual Clustering Paradigm for Resource Management in Cognitive Femtocell Networks

Faisal Tariq, Laurence S. Dooley, Adrian S. Poulton

Next Generation Multimedia Research Group, Department of Communication and Systems

The Open University, Milton Keynes, UK

Email: f.tariq|l.s.dooley|a.s.poulton@open.ac.uk

**Abstract:**

*Femtocells represent a promising alternative solution for high quality wireless access in indoor scenarios where conventional cellular system coverage can be poor. Femtocell access points (FAP) are normally randomly deployed by the end user, so only post deployment network planning is possible. Furthermore, this uncoordinated deployment creates the potential for severe interference to co-located femtocells, especially in dense deployments. This paper presents a new femtocell network architecture using a generalized virtual cluster femtocell (GVCF) paradigm, which groups together FAP, which are allocated to the same femtocell gateway (FGW), into logical clusters. This guarantees severely interfering and overlapping femtocells are assigned to different clusters, and since each cluster operates on a different band of frequencies, the corresponding virtual cluster controller only has to manage its own FAP members, so the overall system complexity is low. The performance of the GVCF algorithm is analysed from both a resource availability and cluster number perspective, and a novel strategy is proposed for dynamically adapting these to network environment changes, while upholding quality-of-service requirements. Simulation results conclusively corroborate the superior performance of the GVCF model in interference mitigation, particularly in high density FAP scenarios.*

## 1 Introduction:

The impetus for the rapidly increasing demand for high data-rate mobile support in indoor environments has been the inexorable shift in users' focus to more data intensive applications like network video gaming, video conferencing, internet-based TV and video-on-demand. Studies reveal that up to two thirds of all mobile data demand is now generated from indoor scenarios [1]. The provision of high indoor data rates is difficult for conventional cellular services since high wall penetration losses lead to attenuated received signal strengths. The problem is particularly severe in the cell-edge areas furthest from the serving macrocell *base stations* (BS). Increasing the BS transmit power is not a viable solution to this problem as it concomitantly increases the co-channel interference to users located at the edge of neighbouring cells. Similarly, deploying more BS is not feasible due to the probative costs involved.

*Femtocell access points* (FAP) are an emerging technology specifically designed to address these challenges. They are inexpensive, plug and play devices which provide high data rates in indoor scenarios [2]. The typical radio range for a home femtocell is <10m, which means the transmit power requirement is significantly lower than a conventional BS. When located indoors, *mobile stations* (MS) connect to the FAP instead of the macrocell BS, with all network traffic being backhauled via either a wired xDSL (digital subscriber line) or fibre optic network.

The short distances involved mean the transmitter-receiver link is robust and the received *signal-to-interference noise ratio* (SINR) is sufficiently high to enable higher-order modulations to be used to secure higher throughputs. Other advantages which FAP technology affords include the MS incurring significantly low power to connect to the FAP, thus prolonging battery lifetimes and the low power transmissions enable the spectrum to be reused more often so system capacity per unit area is increased.

While FAP is an attractive solution, there are many technical, regulatory and economic obstacles to be overcome to achieve successful deployment [3]. Due to the anticipated high operational densities, managing interference is the most significant challenge for successful femtocell deployment. There are a plethora of interference management techniques including interference cancellation [4], interference randomization [5] and interference avoidance or coordination [6]. Interference cancellation for instance, requires multi-antennae and a hardware intensive signal processing capability at the receiver, while randomization averages the interference on user equipment by randomly hopping between channels. Although this latter approach performs better in certain environments, the improvement is limited compared to avoidance or coordination techniques, which have become the preferred solution since they do not increase the complexity of the transceiver system.

Interference management between macro and femtocells and also among femtocells requires some form of collaboration. The macro and femto tiers share both resource allocation and radio environment information, which can be exploited to attain lower interference [7]. Power control techniques have been widely used for interference reduction, where both the macro and femtocell transmission powers are minimised while maintaining the *quality-of-service* (QoS) provision. In large-scale deployments, a distributed model [8] is more expedient compared to a centralised solution, with game theoretic approaches affording both cooperative and non-cooperative methods to determine the optimal Nash equilibrium power value for a given set of objectives and constraints [9], [10], [11]. *Reinforced learning* (RL) is a promising alternative technique for interference minimisation, where the FAP combines the current transmission experience with previous experiences to decide the most appropriate channel and power level for the next transmission [12], [13], [14]. In rapidly varying radio environments however, RL can hinder the rate of convergence to a quiescent state and the FAP training period can become too long, resulting in system inefficiencies.

In *cognitive radio* based approaches [15], [16], SINR measurements are made for all femtocell and macrocell users and fed back to the BS, which then decides the most appropriate operating channel, the transmission power and SINR distribution at any location to ensure either minimal or no interference occurs. Improvements to these techniques are restricted however, by the power required for sensing, the accuracy of channel state information and redundant data incurred for real-time information sharing.

From a femtocell resource management perspective, with the likelihood of progressively larger scale deployments, a centralised model is going to become intractable because it is not scalable and incurs high redundant control data overheads [17]. Conversely, a fully distributed management model may lead to undesirable situations where a unilaterally poor choice for one FAP, particularly in dense deployments, can have serious implications for several other FAPs. Moreover, as deployment densities and radio environments vary dynamically, the resource management architecture must be adaptive to different scenarios. This means hybrid management paradigms [18] are a more propitious option for femtocell networks, with some functionality controlled centrally and some devolved to either the FAP or a local *femtocell gateway* (FGW), which is an intermediary between the FAPs and the *radio network controller* (RNC) in undertaking certain control and management functions.

Since femtocells are arbitrarily positioned by the end users, existing radio resource management paradigms [19], which have been designed for pre-deployment resource planning in either macro-cellular or ad-hoc wireless networks, are simply not applicable in a femtocell context. The absence of any coordination between femtocells means the coverage of one femtocell can overlap with another, resulting in harmful interference [20]. This problem is compounded in high density deployments, so effective interference management strategies are essential in facilitating successful femtocell network operation.

The corollary is that the underlying resource management architecture needs to be redesigned to incorporate the distinctive features of femtocell technology. The resource manager for example, must be flexible and adaptive to accommodate sudden changes in the radio environment, network topology and resource availability. This provided the motivation for the new *generalised* resource management paradigm presented in this paper, which does not require any change in its physical architecture. Introducing the concept of FAP clustering, femtocells are classified into *virtual clusters*, with this terminology reflecting that cluster members are logically linked together, as opposed to the traditional ad hoc clustering sense, where members are located within some defined distance of a *clusterhead*. Members of a virtual cluster may not necessarily be physically co-located, but instead are grouped together to exploit the same set of channels according to a minimum interference generation criterion.

The initial idea of logical clustering was introduced in [21], with a *virtual cluster formation* (VCF) algorithm being applied to a rigid clustering framework, and FAP location information used to create the respective virtual (logical) FAP clusters. The VCF algorithm maximises the minimum distance between the FAPs of any cluster, thereby minimising the overall interference. Since the location of the FAPs remains relatively constant, the corresponding computational complexity of this minimax clustering solution is low compared to other available interference minimisation methods. In this paper, a complete *generalised virtual clustering femtocell* (GVCF) framework is formulated and analysed, and its performance is rigorously evaluated in various network scenarios. The new GVCF paradigm offers much greater flexibility as it can automatically adapt to changes in both the available resources and radio environment, seamlessly handling situations such as when users either leave or join the network. It also negotiates with the central controller when more resources are required to ensure the requisite QoS is upheld. The fixed cluster structure in [21] is relaxed in the GVCF model by regular monitoring and performance evaluation and its ability to adapt the cluster number and their respective FAP members in accordance with fluctuations in both the available resources and a prescribed set of network design constraints, such as the minimum throughput requirement or the maximum transmit power. The corresponding results analysis corroborates the enhanced interference management performance and adaptive capability of the new GVCF model in various network scenarios, especially high density deployments.

The remainder of the paper is organised as follows: Section 2 describes the two-tier system model including the various interference options, a cross-tier interference minimisation technique, and both the path-loss and channel allocation models. Section 3 explains the new virtual clustering architecture, particularly the GVCF framework, while Section 4 rigorously analyses the performance of the GVCF paradigm in terms of throughput, received SINR, and the average reuse distance for varying FAP deployment densities and channel availabilities. Finally, Section 5 presents some concluding comments.

## 2    System Model

A dual-tier (macro-femto) tier system model is considered in this paper, where the femtocell is overlaid upon the macrocell system. The macrocell is hexagonally-shaped, with a BS located at the centre of the cell, and comprising three sectors with MSs being uniformly distributed in each sector. In contrast, femtocells are circular in shape and uniformly distributed inside every macrocell. The FAP is assumed to be located at the centre of a femtocell and the MSs connected to it are uniformly distributed across the femtocell. FAP transmission power is fixed, though different power levels are used for the inner and outer cell MSs. A closed access mechanism is adopted for all femtocells so only authorised MSs can connect to a particular FAP. Macro and femtocell coexistence is assumed, with both operating in the same 10MHz spectrum, which is divided into 180KHz wide equi-spaced channels in an analogous way to the 3GPP LTE definition [22].

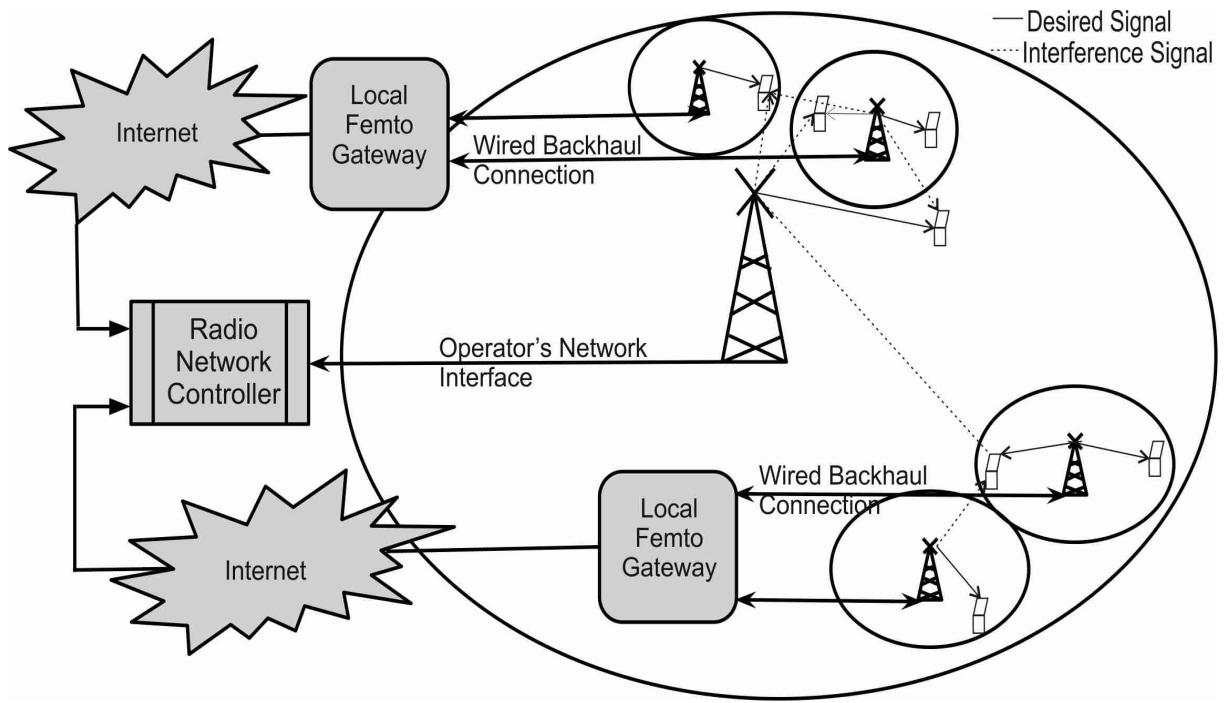

**Figure 1: Example joint macro-femto deployment arrangement with backhaul network**

The interconnection network for joint macro-femto deployment, including the downlink (DL) interference scenarios is shown in Figure 1. The FAPs are connected to a local *FGW* which retains some FAP control functionality relating to the registering of FAP and its user, assisting in the initial cluster configuration, allocating available femto-tier resources, managing local disputes, routing traffic in both directions and most importantly as a link between the RNC and FAPs. The FGW maintains macro-layer communications via the RNC, the internet and the operator's interface (also known as the X2 interface), which routes traffic in harmony with the RNC. Since the FGW plays a crucial role in cross-tier information sharing acting as an intermediary between the macro and femto tiers, the virtual clusters are formed in the FGW.

Importantly at any location, the macro and femtocells use mutually exclusive sub-channels to avoid macro-to-femto and femto-to-macro interference. This is achieved by employing *fractional frequency reuse* (FFR), in which sub-channels used by a MS connected to a specific BS in one sector are allocated to a MS connected to a FAP in a different sector [23], [24]. The new GVCF paradigm addresses the problem of both cross and co-tier interference management in the DL, with special emphasis on femto-to-femto interference minimisation. A variant of FFR called dynamic FFR is adopted in the GVCF framework for resource sharing and this will be delineated in the next section along with the path loss model used to determine the FAP distance between FAPs able to reuse the same channels, and the MS channel allocation scheme.

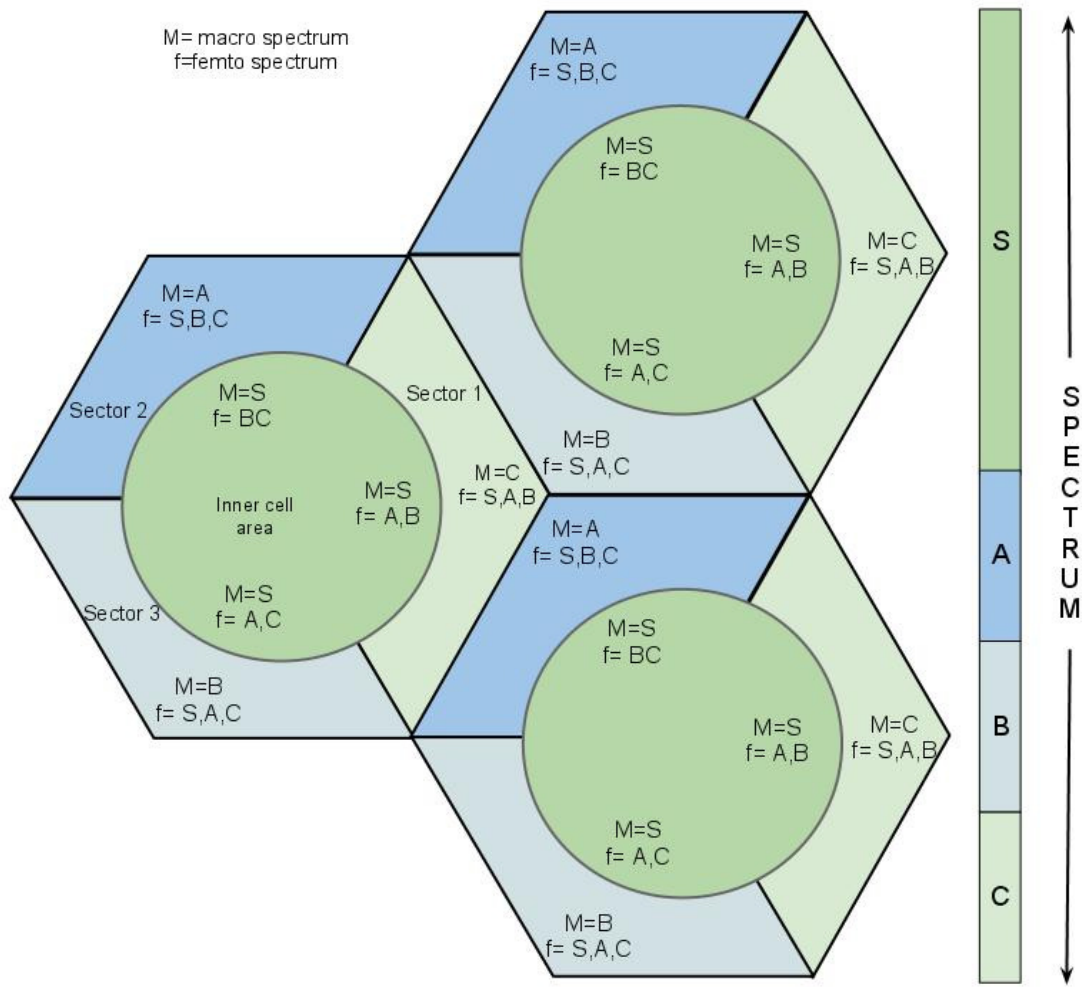

Figure 2: FFR-based resource sharing in joint macro-femto deployments.

***Dynamic FFR:*** Figure 2 illustrates the FFR resource sharing scheme for a macro and femtocell network with 3 macrocells. It divides each macrocell into an inner and outer area, with the latter being further sub-divided into multiple sectors. In dynamic FFR [25], the number of channels available in each sector varies dynamically in proportion to the number of user in the area, with the femto layer being subsequently updated about any changes on a regular interval. Each colour in Figure 2 represents an available spectrum band for the macrocell users in that area, with femtocells being prohibited from using these frequencies.

For example, in the outer cell area of sector 1, macro users can use channel C, so femtocells are only allowed to use the other three channels in this area, namely S, A and B. Vigilance is required however, for femtocells located at the border of a cell as they may interfere with the macro users in an adjacent macrocell. Likewise in the outer cell area of sector 3, femtocells can use the entire spectrum except channel B. For the inner cell areas, macrocell users are allowed to use the S channels, so femtocells are not allowed to use S channels in this area. In addition, to avoid the S channels, femtocells are barred from using the channels of the macrocell users in the outer cell in that sector.

*Path Loss Model:* For indoor scenarios, the WINNER II [26] path loss channel model is utilised, with (1) and (2) giving the indoor *path losses* (PL) for the *line-of-sight* (LOS) and *non-line-of-sight* (NLOS) cases respectively. In the NLOS situation, when there are walls between the transmitter and receiver, an additional *wall penetration loss* ($L_{WP}$) component is included:

$$PL_{LOS}^f = 18.7\log(d) + 46.8 + 20\log\left(\frac{f_c}{5}\right) \quad (1)$$

$$PL_{NLOS}^f = 20\log(d) + 46.4 + 20\log\left(\frac{f_c}{5}\right) + L_{WP} \quad (2)$$

where $d$ is the distance of the user from the FAP in metres, $f_c$ is the carrier frequency, $L_{WP}$ is the wall penetration loss (dB), and $PL_{LOS}^f$ and $PL_{NLOS}^f$ are the femtocell path losses (dB) for the LOS and NLOS signals respectively.

Wall penetration losses vary according to such factors as: transmission frequency, angle of arrival of the signal, thickness of the wall and the material used in the wall, so to simplify the calculations, $L_{WP}$ = 5dB and $L_{WP}$ = 10dB are respectively considered as the internal and external wall penetration losses. Since MSs connected to femtocells are usually located inside a building and so only encounter relatively thin internal walls, $L_{WP}$ = 5dB is applied for received signal power calculations. Conversely, as interference generated by femtocells located in other buildings or houses has to pass through at least two external walls to reach the MSs connected to femtocell, for these interference power calculations, $L_{WP}$ = 10dB is used for each wall loss, resulting in a total penetration loss of 20dB.

*Channel Allocation:* The GVCF model assumes every FAP is responsible for allocating channels to its member MSs, based upon feedback from the MSs on the respective received SINR. To assign the best available channel, each FAP calculates the *carrier-to-interference ratio* for all MSs attached to it, which is formally defined as:

$$\frac{C}{I} = \frac{P_{t0} h_0}{\sum_j P_{nj} + N_0} \quad \ldots \quad \ldots \quad (3)$$

where $C$ is the carrier power, $I$ the interference-plus-noise power, $P_{t0}$ is the transmit signal power, $h_0$ the channel power gain, $P_{nj}$ the received interference power on channel *n* from user *j* and $N_0$ is the noise power.

## 3 The GVCF Paradigm

### 3.1 Logical Clustering Architecture

Clustering has been widely investigated in both the wireless sensor and ad hoc network domains [27], with the normal approach being to select a *clusterhead* from a group of nodes according to some criterion. Neighbouring nodes are then assigned membership of a cluster based upon for instance, being physically co-located. In contrast, the GVCF model uses virtual (logical) clusters which are based on an interference-based Euclidean distance measure.

The principal motivation behind the development of the GVCF paradigm is the key challenge in femtocell networks of interference management. Virtual clusters are formed using a minimax criterion by combining FAPs under a *virtual cluster controller* (VCC) in which all the FAPs operate on the same set of channels, while concomitantly maximising the closest FAP distance. The rationale for the GVCF model is that as power exponentially decays with distance, the FAP furthest away from a particular FAP will correspondingly generate the lowest interference. The corollary being that by maximising the distance of the closest FAP operating on the same channels, the interference is correspondingly minimised.

Figure 3 shows the block diagram of the logical architecture of the GVCF virtual clustering femtocell network. Based on the FFR distribution and the latest usage information acquired from the macro layer, the RNC firstly informs the FGW about the channel set available for allocation in a given area. When FAPs are switched on, they automatically register with the RNC via the FGW and then depending on the highest number of users connected to a FAP, the FGW determines the number of VCC in accordance with the following relationship:

$$N_{VC} = \frac{N_{Ch}(k)}{max(N_f)} \quad \quad \quad \quad \quad \quad \quad \quad (4)$$

where $N_{VC}$ is the number of virtual clusters, $N_{Ch}(k)$ is the number of channels available in the $k^{th}$ area and $N_f$ is the number of femtocell users. The system continually monitors the performance and identifies network variations such as, changes in resource availability or in the radio environment like when a user leaves the network. In these circumstances, (4) is recalculated and the VCCs are reconstructed as will be elucidated fully in Section 3.2.

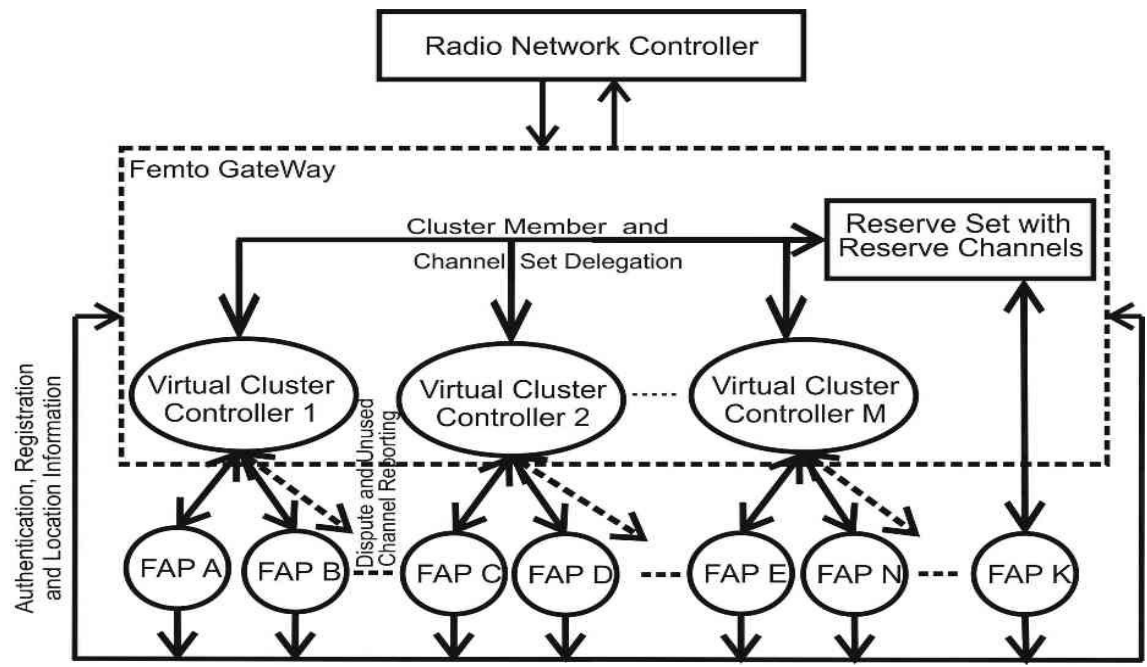

**Figure 3: Logical diagram of the *generalized virtual clustering femtocell* (GVCF) network system.**

Each VCC has specific resource allocation functionality such as, assigning the channel set to cluster member FAPs and managing MS disputes on behalf of its FAP membership, though in the unlikely event of a dispute occurring with either a macrocell user or a MSs connected to a FAP belonging to another VCC, then it is forwarded to the FGW for arbitration. The FGW cooperates with the macrocell BS via the RNC to create a list of channels available for allocation in a certain area using the dynamic FFR technique described in Section 2. Unlike distributed resource allocation approaches, where each FAP independently chooses a channel, the new virtual clustering architecture devolves this task to the VCC which maintains an updated list of available channels. Furthermore, distinct from centralised resource management where every decision, including channel assignment, is performed by the RNC, each VCC takes responsibility for channel set allocation and dispute management on behalf its cluster members. This means the GCVF model inherently provides hybrid resource management, combining the best features of the centralised and distributed resource allocation models. It also saves a significant number of redundant data transfers between each FAP and the FGW and/or RNC.

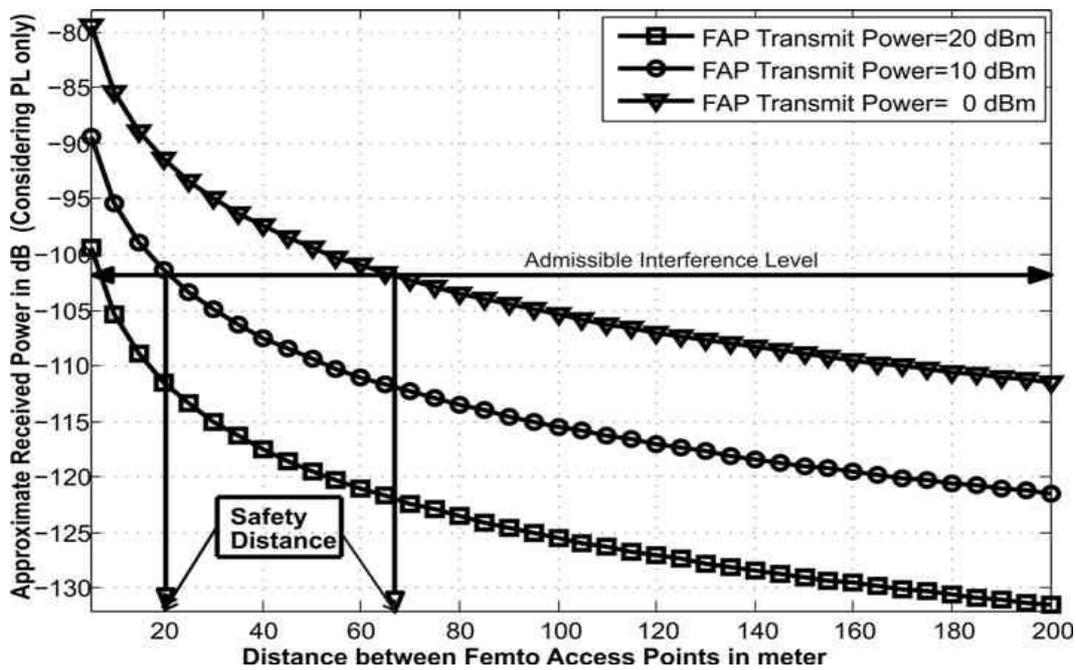

**Figure 4: Safety distance $D_{th}$ measurements for FAP deployments at different transmission powers**

Since the FAP is connected by a wired network, its approximate location is known by the FGW. In addition, femtocell positions can be detected and the network topology constructed to a certain level of accuracy using RF and time difference of arrival [28], [29] measurements, so position information coupled with RF measurements from the respective MSs can be exploited to obtain an accurate FAP location. The GVCF algorithm then assigns each FAP to a designated VCC, which provides access to a set of channels.

GVCF creates and maintains a reserve channel list $Ch_R$ derived from the reporting of unused channels by the FAPs. This list is periodically updated via the FGW, and also includes unused macrocell BS channels informed by the RNC. These reserve channels are allocated to

FAPs either in the case of disputes or to members of the reserve set $S_R$, which includes those FAPs that failed to uphold the *safety distance $D_{th}$*. This is the distance all FAPs must sustain from their co-channel FAP to ensure effective femtocell operation, and is determined by setting the maximum level of admissible interference and then calculating the corresponding minimum distance requirement to preserve the SINR level. An estimate of $D_{th}$ can be obtained from the PL model defined in (1).

Figure 4 illustrates how the safety distance threshold $D_{th}$ is determined. The graph shows the approximate interference power received by a user from a neighboring FAPs at various distances and transmit powers (0, 10 and 20dBm), while maintaining the FAP transmit power to which the user is connected constant at 10dBm. The plot considers only the PL components as this represents the worst case scenario. Actual interference will be lower when cognisance of the fading and shadowing components is made. Depending on the threshold set for the system, the safety distance can vary significantly, so in this paper $D_{th}$ ~-102dB which corresponds to a distance of 20m, which is sufficient to ensure there are no overlapping FAPs, which can lead to severe interference when they are operating on the same frequency. The GVCF algorithm will now be described in detail.

## 3.2 Generalised Virtual Cluster Framework (GVCF)

The flowchart of the complete GVCF algorithm is shown in Figure 5. All the key system parameters are firstly initialised including: the number of FAPs and MS; the safety distance $D_{th}$ and the number of available channels for the area under consideration. It also creates the reserve set $S_R$, and determines the number of virtual clusters ($N_{VC}$) and the inter-FAP distance matrix for *N* FAPs, which is given by:

$$D = \begin{bmatrix} d_{11} & d_{12} & \cdots & d_{1N} \\ d_{21} & d_{22} & \cdots & d_{2N} \\ \vdots & \vdots & \ddots & \vdots \\ d_{N1} & d_{N2} & \cdots & d_{NN} \end{bmatrix} \quad (5)$$

where $d_{ij}$ is the Euclidian distance between FAP$_i$ and FAP$_j$

$$d_{ij} = \sqrt{(x_{FAP(i)} - x_{FAP(j)})^2 + (y_{FAP(i)} - y_{FAP(j)})^2} \quad (6)$$

and where $x$ and $y$ are the Cartesian co-ordinates for each of the $N$ FAPs and $d_{ij} = d_{ji}$. Note, since all $D$ diagonal elements are zero i.e., $d_{ii} = 0$ for $i=1,2,....N$, these are excluded from the minimum distance calculations.

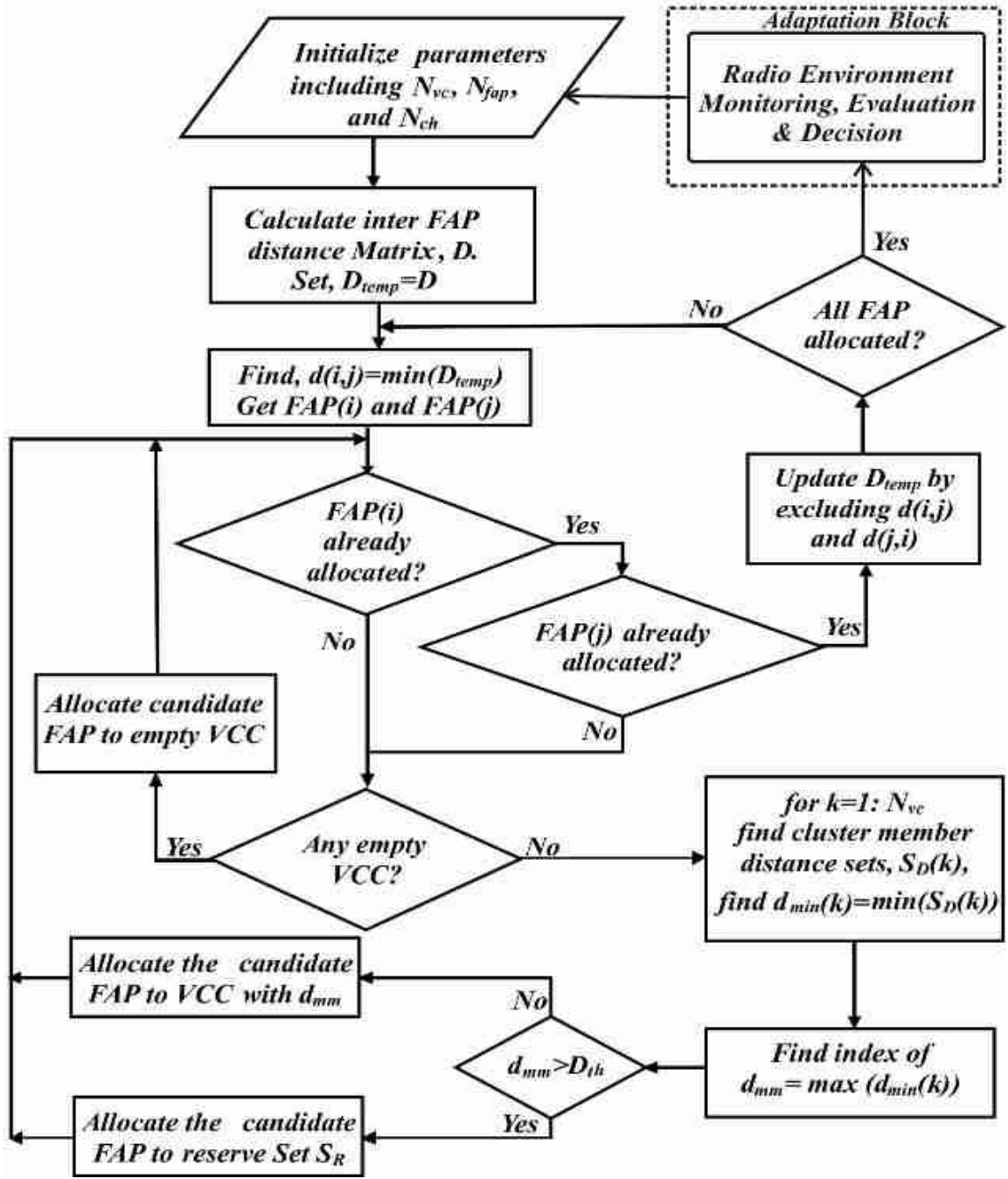

**Figure 5: Flowchart for the *generalised virtual clustering framework* (GVCF) algorithm.**

Following initialisation, GVCF identifies the FAP pair from $D$ with the minimum Euclidean distance. The distance matrix $D_{temp}$, contains all the unallocated FAPs, so when no FAP is allocated $D=D_{temp}$. For each candidate FAP, GVCF firstly checks whether the FAP has already been allocated. If it has not, the algorithm ascertains whether there are any empty VCC and if so, the FAP is duly allocated to a vacant VCC. If both FAPs have already been assigned, then the distance pair is excluded from $D_{temp}$ and both are expunged from the unallocated FAP list. The next closest pair of FAPs from $D$ is then sought and the above process repeated. When every VCC has at least one member, then for any further FAP allocation, the VCF algorithm chooses the distance from $D$ of the candidate FAP to all members of the VCCs. From these inter-FAP distances, the minimum value to each VCC is selected and the FAP is assigned to the VCC which has the highest minimum distance $d_{mm}$, to all the other FAPs belonging to that VCC, subject to the safety distance $D_{th}$ being maintained If the FAP cannot uphold $D_{th}$, it is assigned to the reserve set $S_R$ instead, whose members are allocated reserved channels. $D_{temp}$ is then updated, with the FAP pair assigned during this iteration being excluded, and the procedure repeated until all FAPs have been allocated.

After all the FAPs are allocated, the GVCF algorithm begins the adaption phase by continually monitoring and evaluating the performance of the current clustering arrangement as highlighted in the box in Figure 5. For a given constraint such as, an application specific data-rate requirement, if the existing cluster arrangement cannot uphold the requisite performance, then the RNC is requested to allocate more channels in order to increase the number of clusters. Upon receiving these, the iterative virtual clustering process is repeated. This adaption mechanism importantly identifies radio environment changes such as a MS or FAP either joining or leaving the femtocell network, with the clustering algorithm adjusting accordingly the cluster number and reassigns FAPs to other VCCs to either improve or sustain performance. This uniquely affords the GVCF paradigm flexibility in its ability to automatically respond to changing radio environment conditions and unforeseen network situations.

From a computational complexity perspective, the new virtual clustering paradigm is very efficient as it principally involves the processing of FAP coordinates, so the order of time complexity increases linearly with the number of femtocells deployed.

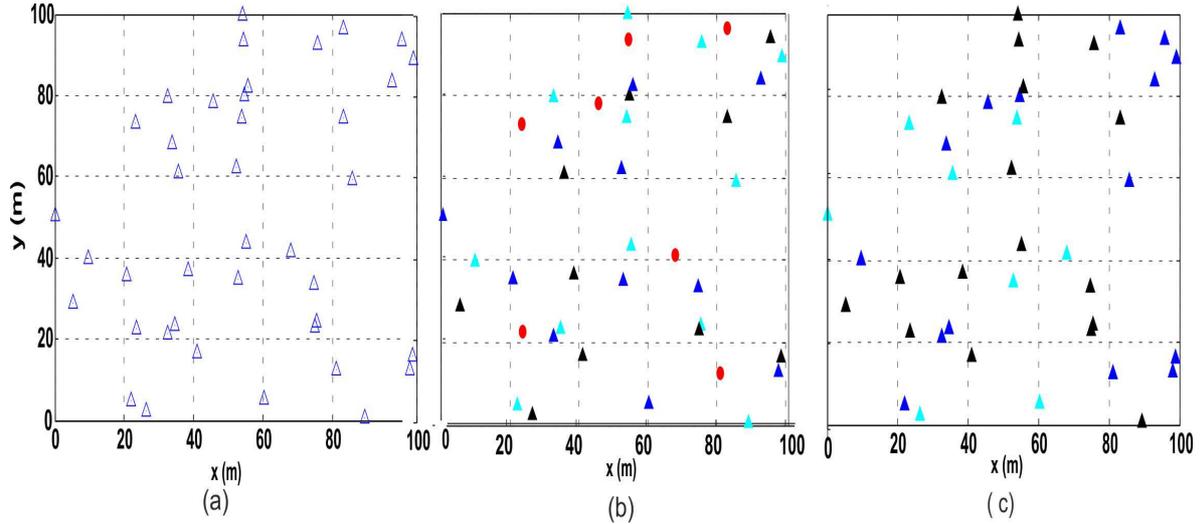

**Figure 6: FAP deployment scenarios: (a) before cluster formation, (b) after clustering (applying GVCF), and (c) the non-clustered solution, where each colour represents the channels of a VCC.**

Figure 6 (a) and (b) respectively show an example FAP distribution both before and after resource allocation is performed using the new GVCF framework, while Figure 6 (c) displays the corresponding random resource allocation. In Figure 6(b), each colour represents a cluster, while in Figure 6(c) each colour represents an equivalent channel set for the non-clustering solution. A comprehensive performance analysis of the virtual clustering model will now be presented.

## 4  Results Analysis

To evaluate the performance of the GVCF paradigm, a 200m squared area of one sector in a hexagonal macrocell was considered, for four specific FAP node deployments of 50, 100, 150 and 200. The number of available channels in the area varied between 4 and 20. As a performance comparator for GVCF, a distributed resource allocation framework was implemented with each FAP able to independently choose its operating spectrum. This scheme is referred to as the *non-clustering system* (NCS) in the ensuing discussion. For the purposes of an equitable comparison, it is assumed the same number of channels is allocated to each FAP by both the GVCF model and NCS. To clarify the nomenclature adopted in this section, the parenthesis values for both GVCF and NCS are the number of virtual clusters or its equivalent, so GVCF (1) represents the worst-case scenario for the new model, with all FAPs operating on a single set of channels so the clustering and non-clustering systems are the same. The simulation test platform was designed and implemented in MATLAB$^{TM}$, with all the various network environment parameters being defined in Table 1.

| System Parameter | Value or Range |
| --- | --- |
| Femtocell radius | 10 m |
| Macrocell radius | 500 m |
| Number of femtocells (FAP) | Various (50, 100, 150, 200) |
| Maximum number of MS per FAP | 4 |
| MS noise figure | 8 dB |
| Internal wall penetration loss | 5 dB |
| External wall penetration loss | 10 dB |
| Shadowing | 6 dB |
| Macrocell transmission power | 46 dB (max) |
| Femtocell transmission power | 10 dBm |
| MS minimum QoS requirement | >0 dB |
| Total bandwidth | 10 MHz |
| Carrier frequency | 2 GHz |
| Channel width | 180 KHz |
| Total number of channels | 50 |
| Number of channels available for the femto-tier in the experimental area | <21 |

**Table 1: Network parameters and their corresponding values used in all the simulations.**

Following parameter initialisation, the GVCF algorithm was analysed under a variety of different deployment scenarios and resource constraints. The first series of experiments sought to evaluate and test the performance and flexibility of the GVCF model. The results in **Figure 7** and **Figure 8** analyse the *cumulative distribution function* (CDF) of the received SINR and the corresponding *spectral efficiency* (SE) for various cluster numbers at a FAP deployment density of 50. Although the number of femtocells is relatively low, the graphs reveal significant SINR gain and a corresponding throughput improvement has been achieved by GVCF compared to the NCS. From (4), the number of virtual clusters increases with the number of channels, so the availability of additional channels means the distance between co-channel FAPs is higher and the corresponding average interference experienced by any femtocell is commensurately lower.

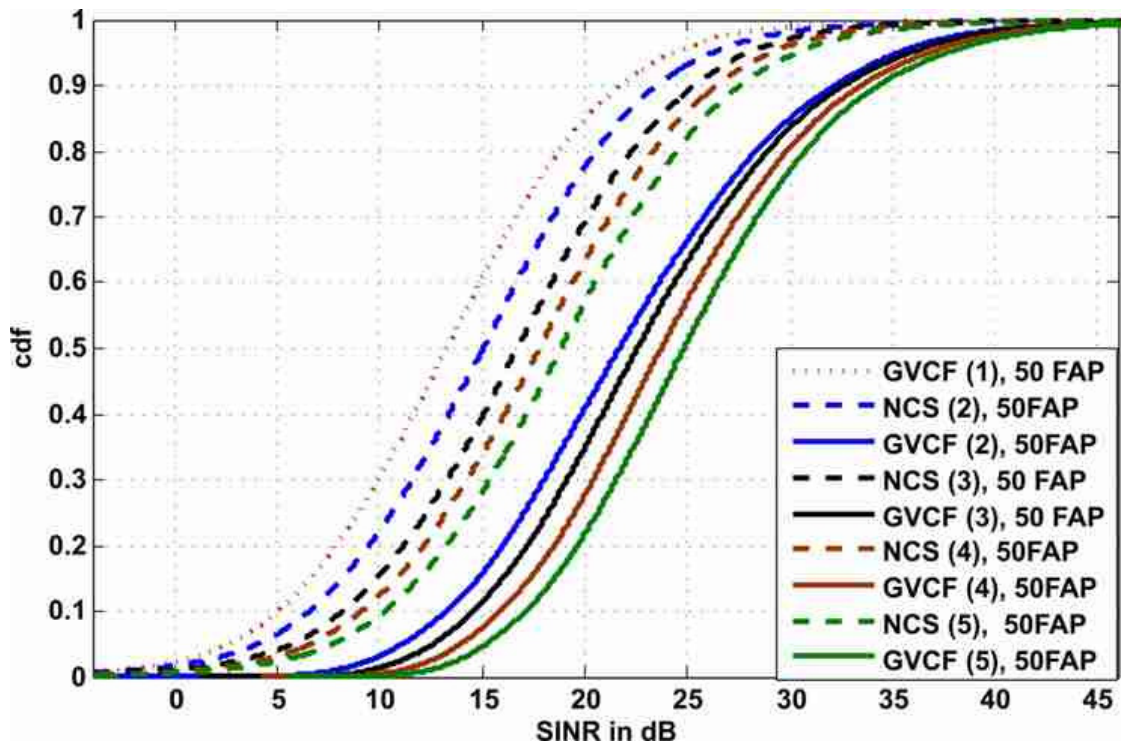

**Figure 7: CDF of received SINR with 50 FAP deployed**

In all cases of channel availability, GVCF outperformed the NCS by a margin of at least 6 to 7dB at the $50^{th}$ percentile value (average received SINR). This improvement is manifest as an average SE gain of approximately 0.5 bps/Hz, and up to 0.8 bps/Hz when the $90^{th}$ percentile value is considered. Percentile values are widely accepted performance metrics by the community [30] and represent the minimum percentage achieved at a particular defined level. Similar performance results are evident at other FAP deployment densities, though as anticipated, both SINR and SE are lower at higher FAP numbers. The principal observation however, is that the GVCF algorithm consistently outperformed the NCS by a margin of at least 5dB, even at the highest FAP density.

The next set of results displayed in **Figure 9** compare the ratio of the number of FAPs which failed to maintain the safety distance threshold $D_{th}$ in the GVCF and NCS models at various FAP deployment densities. When the density is high, i.e. 200, for NCS, almost all FAPs failed to maintain $D_{th}$ even though the channel availability is correspondingly high. With increasing numbers of either VCC or the equivalent number of channels, more FAPs are able to sustain the safety distance threshold, though the level of improvement is significantly better for the GVCF model compared with NCS.

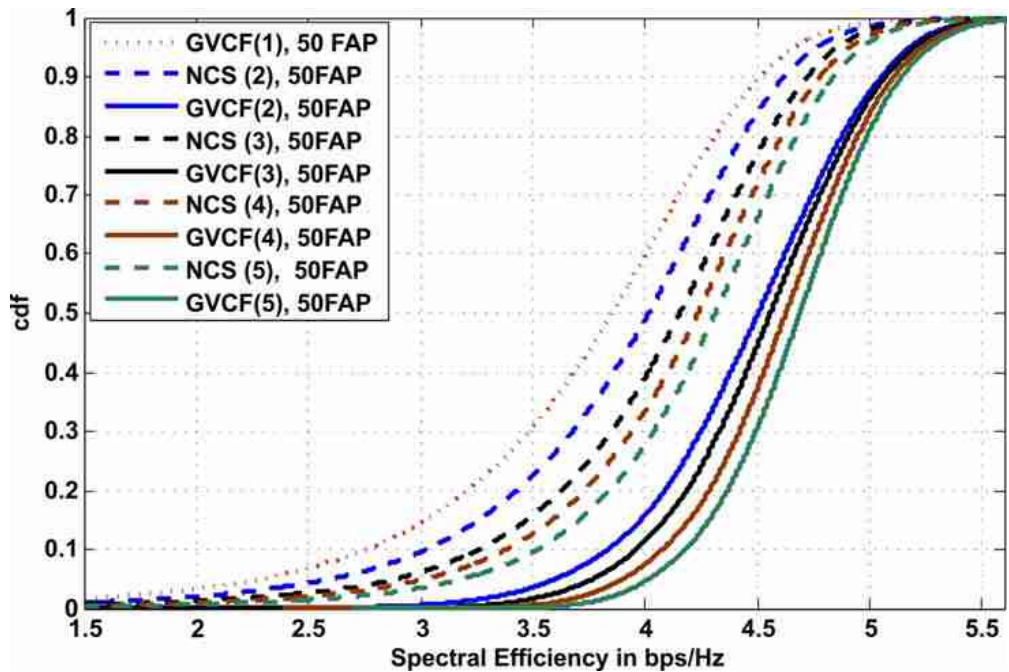

**Figure 8: CDF of the received Spectral Efficiency with 50 FAP deployed**

Intuitively increasing the number of channels will enhance performance, though the main conclusion drawn from these results is that adopting a coordinated virtual clustering strategy provides much better improvement compared to the uncoordinated system (NCS). For example, for a 200 FAP deployment, if 5 clusters are configured, only 10% of the FAPs failed to maintain $D_{th}$ compared with nearly 70% for NCS. This is a significant advance in terms of upholding the minimum safety distance especially in dense femtocell placements. Interestingly at lower deployments for example, 50 FAPs in the given area, only a small improvement is achieved for either two or more clusters, although GVCF is still palpably superior to the NCS solution. This corroborates a key feature of the virtual clustering framework, namely an awareness of either when to demand more channels from the RNC or equally to release extra channels for reuse when they are surplus to requirement. The GVCF paradigm can also crucially determine whether increasing the number of VCC improves the performance, thereby ensuring more efficient usage of the limited available resources.

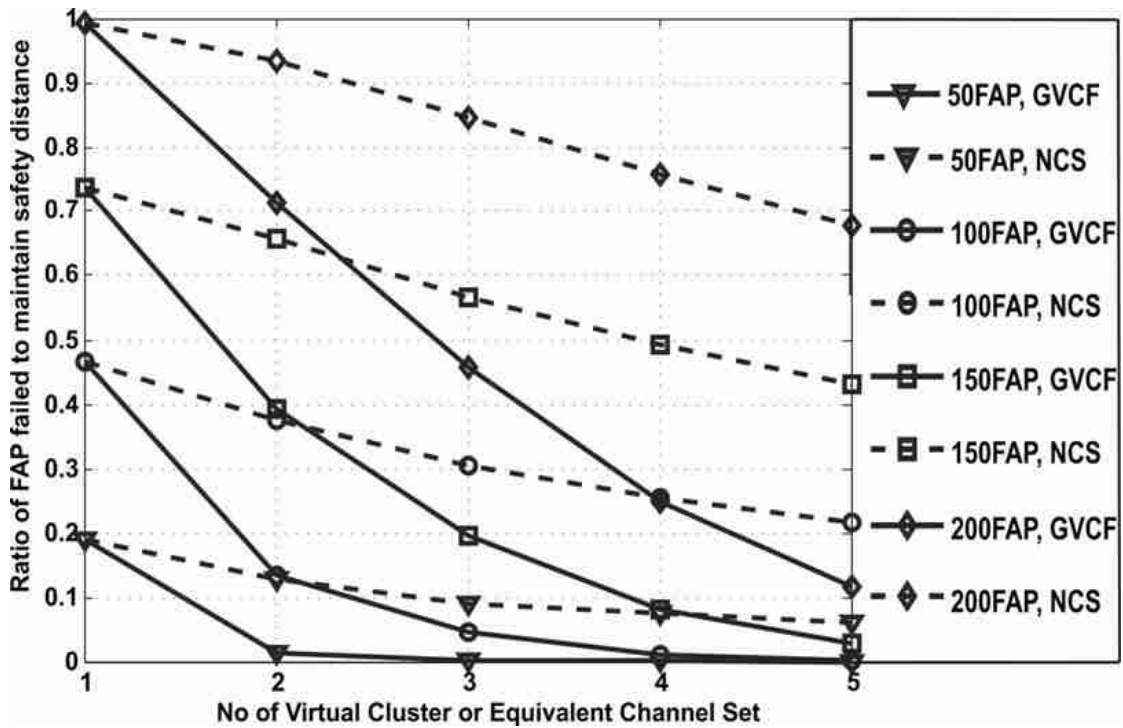

**Figure 9:** The ratio of FAPs failing to maintain the minimum safety distance threshold $D_{th}$ for different number of clusters in both the GVCF and NCS models, with an equivalent channel set.

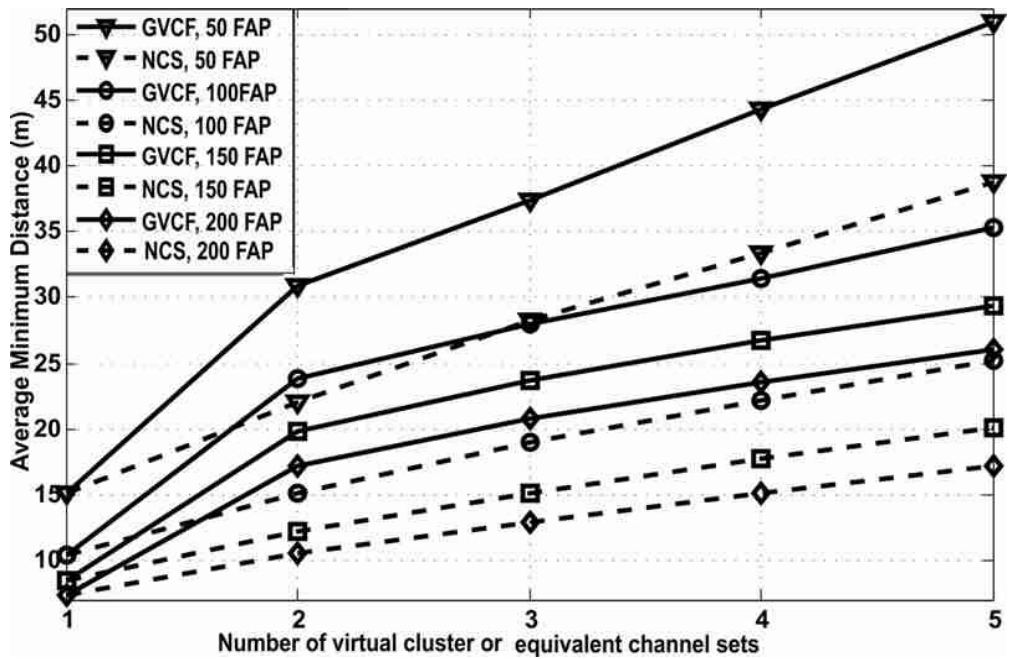

**Figure 10:** Comparison of the average minimum safety distance threshold $D_{th}$ at various FAP densities for different numbers of GVCF clusters and NCS, for an equivalent channel set.

An insightful indicator of system performance is displayed in Figure 10, which shows the minimum safety distance requirement for different cluster numbers and FAP densities. These results were obtained by averaging the distance for either each FAP from the nearest FAP operating on the same channel in case of NCS, or belonging to the same cluster in the case of GVCF for the same scenarios detailed above. The results confirm that with more clusters, the average minimum distance increases and in all cases, GVCF outperformed NCS. While this is an advantageous performance comparison, from a QoS perspective, the SINR and achieved throughput (SE) are the important parameters which reflect system performance and these will now be analysed.

Figure 11 and 12 respectively show the $90^{th}$ percentile SINR and corresponding SE performances at various FAP deployment densities, and cluster numbers or equivalent channel set. This means in 90% of the times the deployed FAPs were able to provide the SINR performance characteristics displayed in the graphs. The overall performance curves for the GVCF model are consistently superior to the NCS at all FAP deployments, and especially at higher densities, i.e., 200 FAPs, where GVCF exhibits a similar performance to that achieved by NCS with only 50 FAPs.

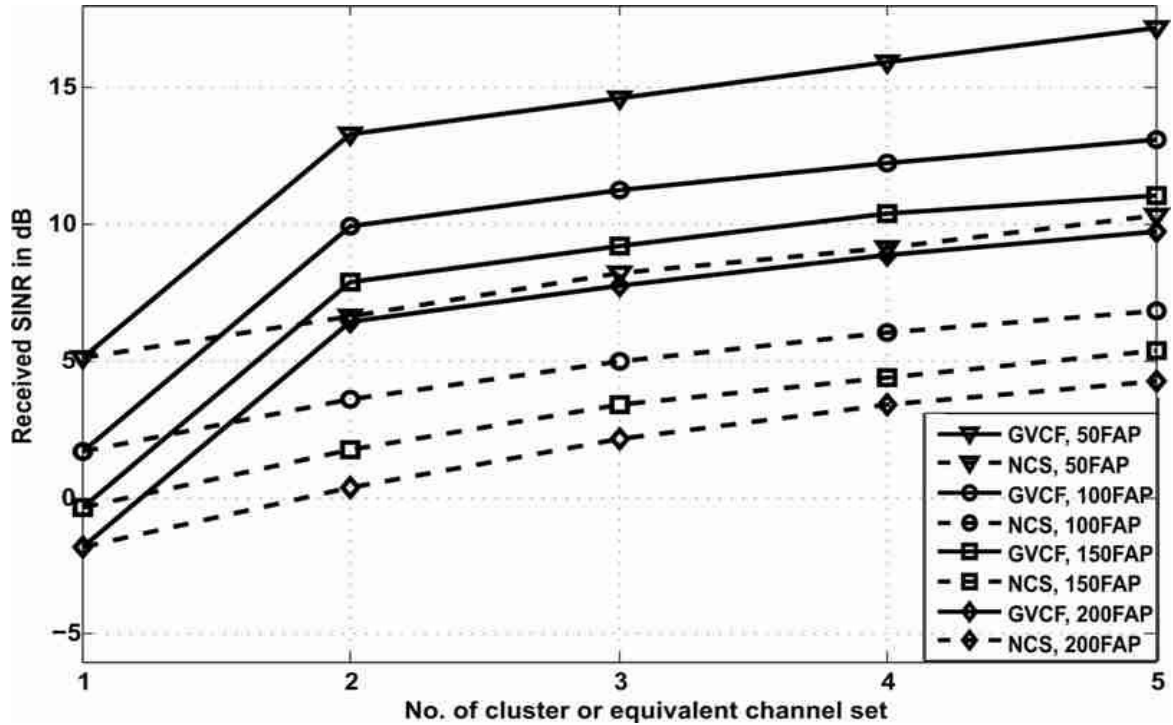

**Figure 11: Comparison between GVCF and NCS for the received SINR ($90^{th}$ percentile) for various FAP densities with different numbers of clusters and an equivalent channel set.**

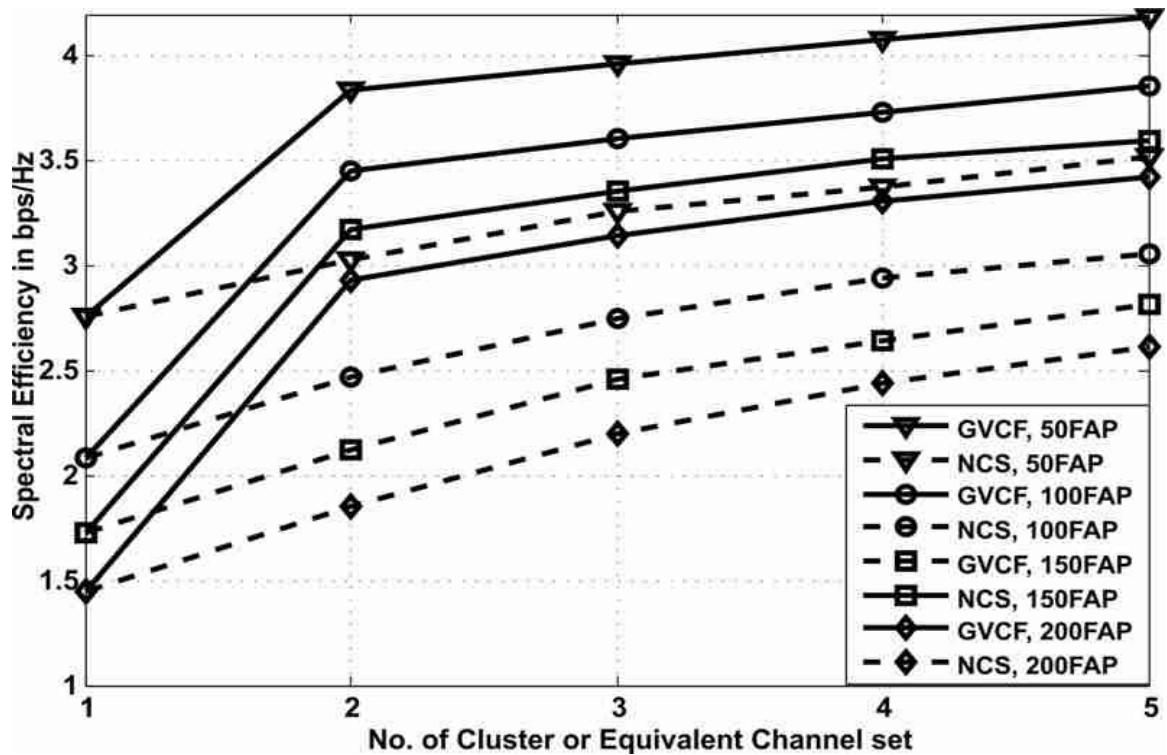

**Figure 12: Comparison of the 90th percentile *Spectral Efficiency* (SE) for various FAP densities with different number of clusters and NCS with an equivalent channel set**

As FAP density increases, the performance of both the clustering and NCS models inevitably degrade, with a corresponding rise in interference. Crucially however, the GVCF is able to uphold an improvement margin of between 6 and 7 dB across all deployment densities. Figure 13 and Figure 14 reveal a similar trend to that observed above for the 50th percentile (average) performance for both systems.

These results provide a valuable awareness into a key characteristic of the GVCF paradigm from a network management perspective. A series of look-up tables (LUT) can be formed for various SINR and SE values and the corresponding resource estimates required for different sets of constraints. This ensures GVCF can uphold a range of diverse QoS requirements, so for example, if there are more than 150 FAPs in a macrocell area and the average bit-rate requirement is 3.6 bps/Hz, then from Figure 14, the femto-tier must have at least 8 channels available in order to form 2 clusters (VCC) to achieve the prescribed QoS as annotated on **Figure 14**. In contrast, the NCS mandates at least 3 sets of channels (12 channels) to achieve analogous performance, so an overall improvement of more than 30% has been achieved by the GVCF system. If exactly the same QoS provision is necessitated at the 90th percentile, then a minimum of 5 clusters (20 channels) are required for the GVCF model, while NCS is simply unable to realise this QoS performance level because it needs more than 5 sets of channels and the maximum channel availability is 20 (see Table 1).

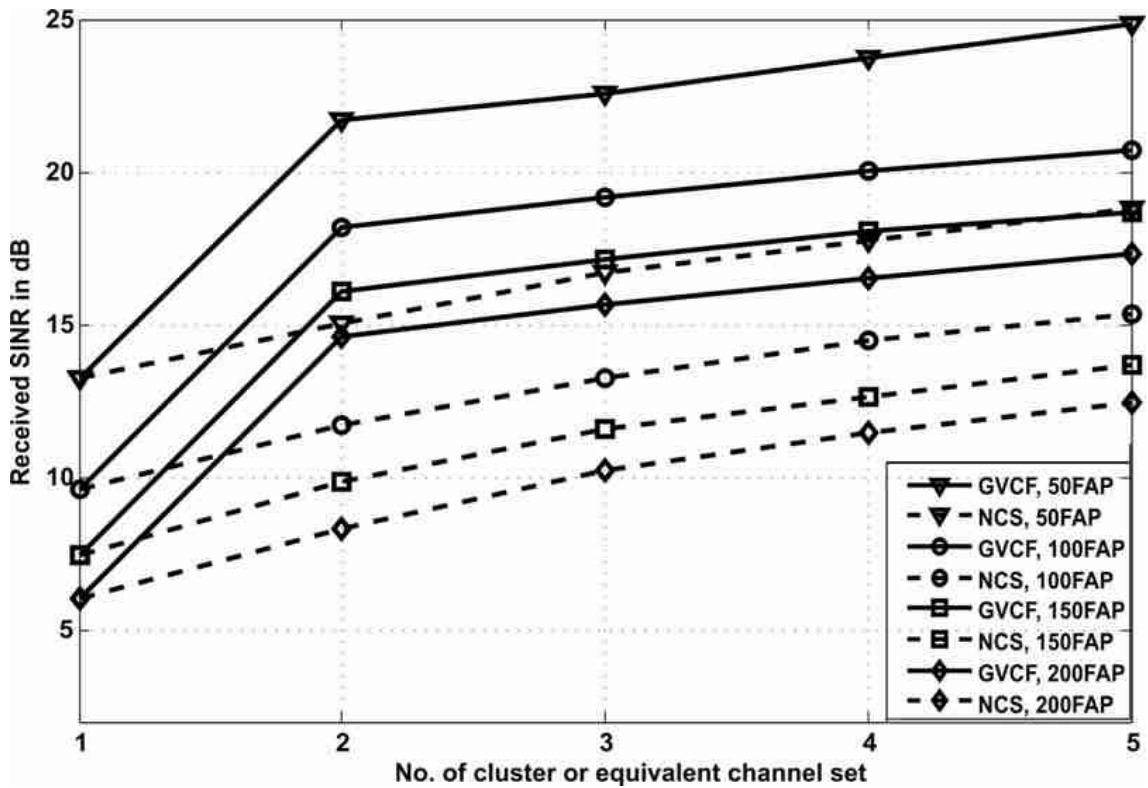

**Figure 13: Comparison of the average received SINR for various FAP densities with different number of clusters and NCS with equivalent channel set.**

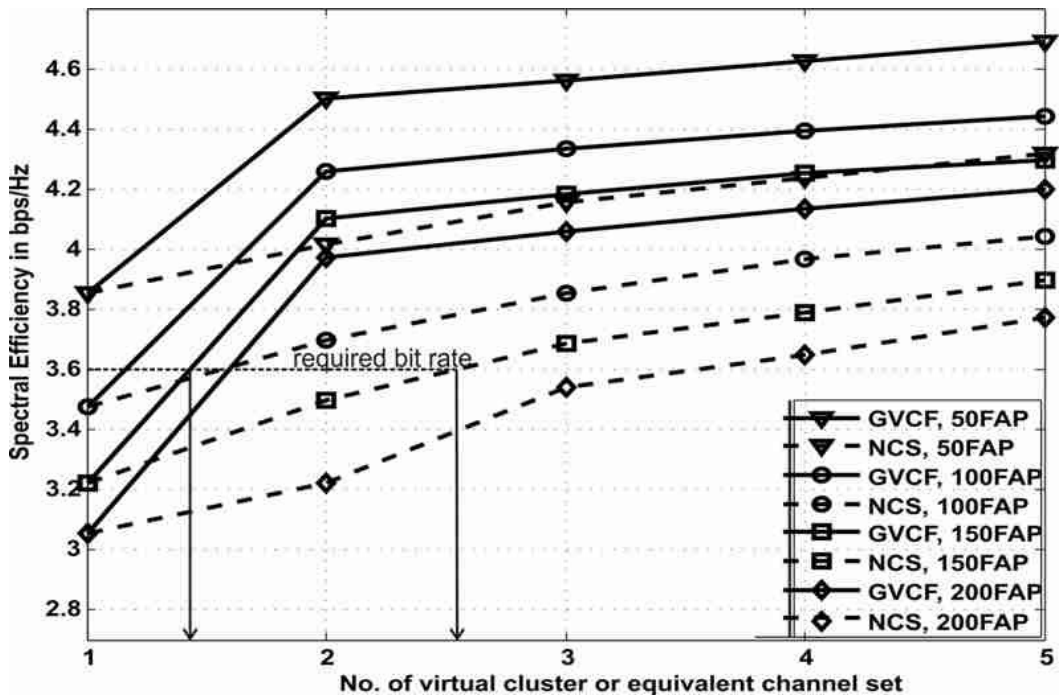

**Figure 14: Comparison of the average SE for various FAP densities with different number of clusters and the NCS with an equivalent channel set**

Conversely, if the maximum number of available channels in the femtocell-tier is 12 and the requisite average SINR is 14dB, then at the 90$^{th}$ percentile, GVCF can service at least 200 FAPs, while for the same QoS requirements, the NCS alternative is unable to serve more than 50 FAP as evidenced in Figure 11. Likewise, if 16 channels are available and there are 150 FAPs, then the maximum achievable SINR at 90$^{th}$ percentile will be 10 dB and in these circumstances, the RNC will need to supply more channels if a better SINR performance is to be accomplished.

Summarising, for a prescribed set of constraints, a LUT can be formed from these performance characteristics so the FGW can communicate with the RNC in order to either demand or release the necessary resources for the femto-tier so the desired QoS provision is upheld. This emphasises the adaptive functionality of the new GVCF paradigm in terms of being malleable to changes in radio environments and network performance by readily adjusting the numbers of clusters and their respective FAP members to always provide a minimised interference solution in comparison with the non-clustered solution.

## 5  Conclusion

This paper has addressed the important dual problems of cross-tier and co-tier interference management in femtocell networks, with emphasis given to the minimisation of intra-femtocell interference in the downlink. A new generalised virtual clustering femtocell (GVCF) architecture for resource management has been presented which employs logical clustering of femto access points (FAP) to achieve interference minimisation and corresponding performance improvements. Simulation results vindicate the rationale for adopting a virtual clustering architecture as it consistently outperforms a distributed random channel allocation system in all network scenarios, providing significant improvements in SINR and throughput, especially at high FAP deployment densities. The inherent low complexity and adaptive nature of the GVCF paradigm allows the number of clusters and their FAP members to be automatically adjusted to either network or radio environment changes such as when a FAP or mobile station either leaves or joins the network.